\documentclass{emulateapj}
\usepackage{apjfonts}
\usepackage{amsmath}

\newcommand{\ledd}{L_{\rm Edd}}

\newcommand{\Mdot}{\dot{M}}
\newcommand{\mdot}{\dot{m}}
\newcommand{\Mdotedd}{\dot{M}_{\rm Edd}}
\newcommand{\etal}{et al.}
\newcommand{\er}{\epsilon_{r}}
\newcommand{\eo}{\epsilon_{r}^{0}}
\newcommand{\tBH}{t_{\rm BH}}

\newcommand{\mdotcrit}{\mdot_{\rm crit}}
\newcommand{\mdotH}{\mdot_{\rm H}}
\newcommand{\EV}[1]{\langle #1 \rangle}

\shorttitle{Low-Efficiency Black Hole Accretion}
\shortauthors{Hopkins, Narayan, \& Hernquist}
\slugcomment{Submitted to ApJ, October 10, 2005}
\begin{document}

\title{How Much Mass do Supermassive Black Holes Eat in their Old Age?}
\author{
Philip F. Hopkins\altaffilmark{1}, 
Ramesh Narayan\altaffilmark{1}, 
\& Lars Hernquist\altaffilmark{1}}
\altaffiltext{1}{Harvard-Smithsonian Center for Astrophysics, 
60 Garden Street, Cambridge, MA 02138, USA}

\begin{abstract}

We consider the distribution of local supermassive black hole
Eddington ratios and accretion rates, accounting for the dependence of
radiative efficiency and bolometric corrections on the accretion rate.
We find that black hole mass growth, both of the integrated mass
density and the masses of most individual objects, must be dominated
by an earlier, radiatively efficient, high accretion rate stage, and
not by the radiatively inefficient low accretion rate phase in which
most local supermassive black holes are currently observed. This
conclusion is particularly true of supermassive black holes in
elliptical host galaxies, as expected if they have undergone merger
activity in the past which would fuel quasar activity and rapid
growth.  We discuss models of the time evolution of accretion rates
and show that they all predict significant mass growth in a prior
radiatively efficient state.  The only way to avoid this conclusion is
through careful fine-tuning of the accretion/quasar timescale to a
value that is inconsistent with observations.  Our results agree with
a wide range of observational inferences drawn from the quasar
luminosity function and X-ray background synthesis models, but our
approach has the virtue of being independent of the modeling of source
populations.  Models in which black holes spend the great majority of
their time in low accretion rate phases are thus completely consistent
both with observations implying mass gain in relatively short, high
accretion rate phases and with the local distribution of accretion
rates.

\end{abstract}


\section{Introduction}
\label{sec:intro}

The growth of quasars and supermassive black holes is a topic of
fundamental interest in cosmology, critically informing theories of
black hole formation and accretion models, and the role of black hole
energetics in the X-ray, UV, and infrared backgrounds and their 
consequences for 
reionization. It is now believed that supermassive black holes
reside at the centers of most if not all galaxies
\citep[e.g.][]{KR95,Richstone98,KG01}.  Furthermore, recent
discoveries of correlations between masses of black holes in nearby
galaxies and either the mass (Magorrian et al. 1998; McLure \& Dunlop
2002; Marconi \& Hunt 2003) or velocity dispersion (i.e. the
$M-\sigma$ relation: Ferrarese \& Merritt 2000; Gebhardt et al. 2000)
of spheroids demonstrate a fundamental link between the growth of
supermassive black holes and galaxy formation.

Observations of the nearby Universe suggest that rapid black hole
growth may be related to gas inflows in the centers of galaxies,
powering intense infrared emission \citep[e.g.][]{Sanders88}, with
mergers or interactions of galaxies providing the gravitational
torques necessary to drive large quantities of gas to galaxy centers
(e.g. Barnes \& Hernquist 1991, 1996) to fuel both starbursts
\citep{MH94,MH96} and rapid, self-regulating black hole growth
\citep{DSH05}.  This picture of quasar activity, whether represented
in analytical or numerical models of the $M-\sigma$ relation
\citep{SR98,Fabian99,CO01,DSH05}, semi-analytical models or
computer simulations of galaxy and
quasar formation \citep{KH00,HM00,DiMatteo03,DiMatteo04,WL03,Granato04}, 
or modeling of
quasar light curves and the quasar and elliptical galaxy luminosity
functions (Hopkins et al.\ 2005a-g), implies that black hole growth
should be dominated by radiatively efficient, relatively short phases
at high accretion rates ($\gtrsim10\%$ of Eddington).

This picture is supported by inferred black hole accretion histories
from the quasar luminosity function
\citep[e.g.][]{Soltan82,Salucci99,YT02,Marconi04,Shankar04,Merloni04a,Merloni04b}
and synthesis models of the X-ray background
\citep[e.g.][]{Comastri95,Gilli99,ERZ02,Ueda03,Cao05,Barger05}, as
well as direct observations of quasar accretion rates at different
redshifts \citep{Vestergaard04,Heckman04,MD04} or models which combine
all of these based on quasar evolution and the observed luminosity
function \citep{H05c,H05e}.  However, these inferences are generally
model dependent, and more importantly are based on quasar observations
which, while in some cases probing X-ray luminosities well below
$L^{\ast}$ and much fainter than bright optically selected quasars, do
not include the (at least locally) very large low accretion rate,
radiatively inefficient population. It is not clear how strongly
constrained is the possibility of black holes gaining a significant amount
of mass in much lower accretion rate phases after periods of early,
bright activity; in these latter phases the radiative efficiency may
drop, and thus the black holes will not usually be observed.

Various authors have attempted to use observational estimates of the
distribution of quasar accretion rates as a function of redshift to
answer the question of whether or not all black holes pass through and
gain their mass in a high-Eddington ratio bright quasar phase
\citep[e.g.,][]{Heckman04,MD04,AGES}.  Although the determination of
the distribution and evolution of quasar accretion rates remains a key
outstanding problem, selection effects limit the questions these
high-redshift observations can answer. These surveys are generally
complete only down to accretion rates $\sim0.1$ of Eddington, and at
any given redshift only a small fraction of black holes are ``active''
with Eddington ratios above this fraction.  Surveys which can probe
lower accretion rates have yielded very different results for the
number and distribution of objects with accretion rates in the range
$\sim0.01-0.1$ Eddington (compare, e.g.\ the estimates of Vanden Berk
et al.\ 2005 from SDSS and Kollmeier et al.\ 2005 from AGES).
Consequently, although the observations described above may suggest
otherwise, these high-redshift measurements of the accretion rate
distribution at large Eddington ratios cannot determine whether or not
a significant population of black holes exists at accretion rates $\sim1\%$
of Eddington, sufficient to dominate their mass growth and total mass
density by $z=0$.

Furthermore, estimating the mass growth of individual black holes (as
opposed to the integrated growth of the population as a whole) in the
observed high-accretion rate phases is not directly possible, as the
duration of this phase is degenerate with the rate at which quasars
are ``triggered'' or ``activated'' at these rates. Even indirectly,
neither quantity is well-constrained, as e.g.\ the duration of this
phase is currently restricted observationally only to the range
$\sim10^{6}-10^{8}$\,yr \citep{Martini04}.

At low accretion rates (relative to Eddington), black holes transition
to radiatively inefficient accretion flows \citep[RIAF/ADAF,
e.g.,][]{NY95,Qua01,Narayan04,XBONGS}, making these objects especially
difficult to observe, and implying that observational surveys may be
an order of magnitude less deep in terms of intrinsic accretion rate
than they are in terms of a radiative ($L/L_{\rm Edd}$) Eddington
ratio.  It is therefore especially important to consider the
distribution of accretion rates at low redshifts, where objects with
low Eddington ratios can be reliably detected, as a constraint on
possible black hole growth in low Eddington ratio states. The fact
that the population of low Eddington ratio sources, which must exist
at all moderate and low redshifts (owing to limits on the integrated
black hole mass and X-ray luminosity density as discussed above), is
observed at $z=0$ to extend to accretion rates as low as $10^{-6}$ of
Eddington strongly emphasizes this point, as these accretion rates are
unlikely to be observable in high-redshift surveys in the near future.

In this paper, we consider the $z=0$ observed distribution of
accretion rates and determine whether or not it allows for significant
mass growth in radiatively inefficient, low accretion rate
states. There are three basic questions one could ask:

{\bf (1)} In which state --- radiatively inefficient, low accretion
rate, or radiatively efficient, high accretion rate --- do black holes
spend most of their {\em time}?

{\bf (2)} In which state do black holes gain most of their {\em mass}?

{\bf (3)} In which state do black holes radiate most of their {\em
luminosity}?

According to current wisdom, the answer to question {\bf 1} is the
radiatively {\it inefficient} state, and that to question {\bf 3} is
the radiatively {\it efficient} state.  However, question {\bf 2} is
rather delicate; it is by no means straightforward to deduce its
answer from the other answers.  The purpose of the present paper is to
investigate question {\bf 2} in depth using the $z=0$ distribution of
black hole masses and accretion rates.

We describe the observed sample from \citet{ho02} and compare with
that of \citet{Marchesini04}, and discuss possible selection effects
in \S~\ref{sec:sample}. In \S~\ref{sec:mdots} we discuss the
dependence of bolometric corrections and 
radiative efficiency on accretion rate and our
accounting for this in determining the distribution of accretion
rates. In \S~\ref{sec:morph} we consider the distribution of accretion rates as
a function of host galaxy morphology.  We discuss several models for
the mass growth of black holes and their implications for the
possibility of mass gain in low accretion rate states in
\S~\ref{sec:models}, and finally summarize our conclusions in
\S~\ref{sec:conclusions}.

\section{The Observed Distribution of Low-Efficiency Accretion Rates}
\label{sec:obs}

\subsection{The Sample and Possible Selection Effects}
\label{sec:sample}

We consider the observed sample of \citet{ho02}, itself an update of
the compilation of black hole masses in \citet{ho99a}, which includes
black hole masses, B-band luminosities, redshifts, and host galaxy
morphological information for 80 nearby supermassive black holes.  The
measurement methods for black hole masses and luminosities, and
possible selection effects, are discussed in detail in \citet{ho02},
but we briefly review them here.

The majority of the black hole masses are derived from spatially
resolved observations of gas and/or stellar kinematics (for a review
of the methods, see e.g.\ Kormendy \& Richstone 1995).  There are a
few exceptions, namely Sgr A$^{\ast}$ in the Milky Way, whose mass is
estimated based on proper motions of individual stars
\citep[e.g.,][]{Sch02,Eckart02,Ghez05}, four galaxies which have
strong water maser emission (NGC 1068, NGC 4945, Circinus, NGC 4258; 
Greenhill et al.\ 1996, 1997, 2000; Miyoshi et al.\ 1995, respectively), and Arp 102B for
which the orbital period of the accretion disk has been determined
\citep{Newman97}.  In \citet{ho02}, the differences between masses
estimated by different means of dynamical modeling of the integrated
spectroscopy (primarily $HST$ spectroscopy of stellar dynamics in the
central regions of each galaxy) are discussed. Three-integral
dynamical modeling of the galaxy surface brightness profiles and
stellar absorption lines is used to estimate the black hole masses,
which tends to give slightly lower black hole masses than the
simplified axisymmetric two-integral modeling of e.g.\ \citet{mag98}
\citep{vdM99,FM00,Gebhardt00}.

For more luminous or distant active galactic nuclei 
(AGN), direct dynamical measurements are no
longer feasible, as the AGN luminosity will overpower the central
stellar emission features. Thus, a large fraction of the sample masses
are determined via reverberation mapping \citep{BM82} of the central
emission lines.  The uncertainties of this method are discussed in
\citet{ho99a,WPM99,Kaspi00,MD01,Krolik01}, but it is generally true
that black hole masses estimated via reverberation mapping agree with
those determined from dynamical estimates to within a factor of
$\sim2-3$. The reverberation-mapped objects of \citet{ho02} are
compiled from \citet{Kaspi00} and \citet{ho99a}, in which the
individual objects are described further.

The optical (B-band) luminosities of the sources in the sample are
determined using the correlation between H$\beta$ luminosity and
B-band absolute magnitude \citep{Yee80,Shuder81}, as calibrated by
\citet{HP01}, which is in principle a more isotropic quantity than
optical continuum luminosity and is measurable in obscured (Type II)
AGN. Where the B-band magnitude is directly available (including all
of the reverberation-mapped sources), it agrees reasonably well with
this determination, with no significant systematic bias
\citep{HP01,ho02}.  We do not consider the 17 PG quasars in the
\citet{ho02} sample, as these objects are chosen from yet another
parent sample, at non-negligible redshifts $z=0.1-0.2$, introducing
stronger selection effects.  Moreover, they are all at reasonably
large Eddington ratios \citep{ho99a,ho02}, and thus are members of the
class of higher-redshift, high-accretion rate radiatively efficient
objects in which we are explicitly not interested.  We have however
checked that including them in our total sample results in a negligible 
difference in the cumulative Eddington ratio distribution and our
subsequent calculations, after normalizing the relative
fraction for the effective volume of the subsample (i.e.\ converting
the observed number to a number density) in
Figure~\ref{fig:mdot.dist} below.  

\begin{figure}
    \epsscale{1.15}
    \centering \plotone{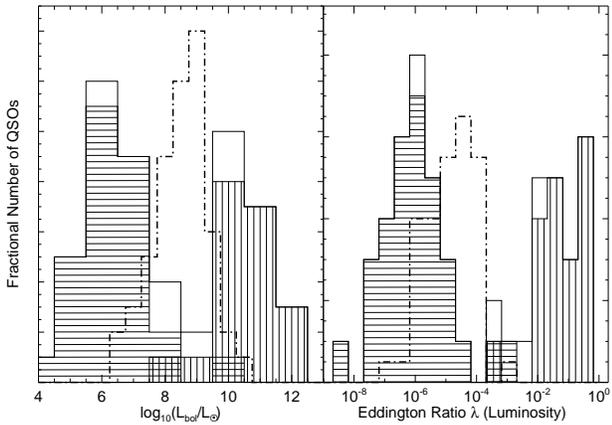} 
    \caption{The distribution of bolometric luminosities (left) and
    radiative Eddington ratios ($\lambda\equiv L/\ledd$, right) in the
    \citet{ho02} sample (solid histograms) and the
    \citet{Marchesini04} radio-selected sample (dot-dashed
    histograms). Open histogram shows the cumulative sample of
    \citet{ho02}, horizontally shaded shows the distribution for
    objects with stellar or gas dynamical mass determinations,
    vertically shaded shows the distribution for objects with
    reverberation mapping mass determinations.  \label{fig:methods}}
\end{figure}

We briefly consider possible selection effects of the different
methods by which the black hole masses in this sample are measured.
As noted above, strong or bright AGN will overwhelm the stellar
emission in the central regions of the observed galaxy and make direct
measurements of stellar kinematics difficult. Therefore, the objects
with masses measured via stellar kinematics may be systematically
biased towards lower nuclear luminosities, and correspondingly lower
Eddington ratios and accretion rates.  However, the
reverberation-mapped objects are expected to have the opposite bias,
as this method of mass measurement requires resolving the AGN emission
lines, and thus may bias the observed sample to higher Eddington
ratios and accretion rates.

Figure~\ref{fig:methods} shows these effects explicitly. Here, the
distribution of bolometric luminosities of our entire adopted
\citet{ho02} sample is shown (open histograms, left panel). Since we
simply wish to show the luminosity range sampled by the observations
we use a constant bolometric correction ($L=17\,L_{B}$), but we adopt
more accurate bolometric corrections in \S~\ref{sec:mdots}.  The right
panel shows the corresponding distribution of radiative Eddington
ratios.  Horizontally shaded histograms show the distribution for
objects with masses determined using gas or stellar dynamics, whereas
vertically shaded histograms show the distribution for objects with
masses determined using reverberation mapping. There is a considerable
gap in the luminosities probed around $10^{9}\,L_{\sun}$, and a
similar gap in Eddington-scaled luminosities around $10^{-4}-10^{-2}$.
Owing to the minimal overlap in the bolometric luminosities covered by
either method, it is difficult to reliably determine whether the
observed gap in the Eddington ratio distribution is real or represents
the combination of the selection effects of the two methods.

To test this, we consider the sample of black hole masses and
bolometric luminosities from \citet{Marchesini04}, shown in both
panels as dot-dashed histograms.  We consider the FR I sample studied
by these authors, based on the 298 radio galaxies and 53 radio loud
quasars of the 3CR catalogue \citep{Spinrad85}. The black hole masses
are estimated from the stellar luminosities of the host galaxy bulges
(except for the bright radio loud quasars, which we do not consider
here for the same reason we neglect the high accretion rate quasars in
the Ho [2002] sample, although again we have checked and find that the
relative number density is not large enough to change our results).
The bolometric luminosities are based on fitting the nuclear optical
luminosities to the light profiles of the observed galaxies.

The advantage of considering the Marchesini et al. (2004) sample is
clear in the left panel of Figure~\ref{fig:methods}, as it covers the
range of bolometric luminosities in which the \citet{ho02} samples
show a deficit, yielding a combined sample with a smooth (non-bimodal)
luminosity distribution. The Eddington ratio distribution for these
objects is shown in the right panel, and although it peaks at a higher
Eddington ratio ($\sim10^{-4}$) than the stellar/gas dynamics-measured
sample of \citet{ho02}, it has a similar qualitative behavior.
Furthermore, this higher peak most likely owes to the higher
luminosity limit in the \citet{Marchesini04} sample.

Although including the \citet{Marchesini04} sample fills the gap in
bolometric luminosity, the gap in Eddington ratio at $10^{-3}-10^{-2}$
is preserved, a point which is discussed in \citet{Marchesini04}.  If
we consider the combined distribution from the \citet{ho02} and
\citet{Marchesini04} samples, the low-Eddington ratio distribution
peaks at roughly $10^{-5}$, more than low enough that our subsequent
results are all qualitatively unchanged regardless of whether or not
we include the objects in the \citet{Marchesini04} sample.  Based on
this, the ``gap'' at Eddington ratios $\sim10^{-2}$ does not appear to
be a selection effect, and we can rely on the \citet{ho02} sample in
our following analysis.

In the following, we generally do not include the sample of
\citet{Marchesini04} except where otherwise specified, because it is
unclear how to correct for possible bias introduced in considering a
radio-loud sample, given the correlation between Eddington ratio and
radio loudness observed in \citet{ho02}. For further discussion of
these distinctions between samples and possible selection effects, we
refer to \citet{ho02,Marchesini04,Jester05}.  


\subsection{Determining the Accretion Rate}
\label{sec:mdots}

The bolometric luminosity, which we denote by $L$, of a black hole
accreting at a rate $\Mdot$ is $L=\er\Mdot c^{2}$, where $\er$ is the
radiative efficiency.  Therefore, the Eddington rate of accretion for
a black hole of mass $M$ is $\Mdotedd=\ledd(M) / \er c^{2}$, where the
Eddington luminosity of a black hole is well-defined,
\begin{equation}
\ledd \approx 3.3\times10^{4}\,L_{\sun} (M/M_{\sun}). 
\end{equation}
Because the Eddington accretion rate depends on the radiative
efficiency $\er$, which itself can depend on the accretion rate, these
definitions are circular unless we define the Eddington accretion rate
with respect to a canonical radiative efficiency $\eo$.  We follow
standard practice (e.g.\ Narayan \& Yi 1995b; Esin, McClintock, \&\
Narayan 1997; Jester 2005, but see also Chen et al.\ 1995; Blandford
\& Begelman 1999; Marchesini et al.\ 2004 who choose $\eo=1$) in
defining
\begin{equation}
\eo=0.1, 
\end{equation}
which corresponds to the typical radiative efficiency expected from
accretion through a standard efficient \citet{SS73} thin disk.  We
then define the Eddington rate as
\begin{equation}
\Mdotedd=\frac{\ledd(M)}{\eo c^{2}}=\frac{M}{t_{S}}, 
\end{equation}
where 
\begin{equation}
t_{S}=4.2\times10^{7}\,{\rm yr}
\end{equation} 
is the Salpeter time, the mass $e$-folding time for a black hole
accreting at the Eddington rate with $\er=0.1$ \citep{Salpeter64}.  We
can further define the dimensionless luminosity Eddington ratio
$\lambda$ and mass accretion rate Eddington ratio $\mdot$ by
\begin{equation}
\lambda\equiv L/\ledd, \qquad \mdot\equiv \Mdot/\Mdotedd. 
\end{equation}
With these definitions, we have 
\begin{equation}
\lambda=\frac{L}{\eo\, M\,c^{2}/t_{S}} =
\mdot\frac{\er(\mdot)}{\eo},
\end{equation}
where we allow explicitly for the possibility that the radiative
efficiency may depend on $\mdot$.

We can also define the black hole growth timescale, which
corresponds to the mass $e$-folding time for growth at a constant
Eddington ratio,
\begin{equation}
\tBH=\frac{M}{\Mdot}=\frac{1}{\mdot}\,t_{S}.
\end{equation} 
Note that this further lets us define a special value of $\mdot$,
\begin{equation}
\mdotH\equiv t_{S}/t_{\rm Hubble}\approx0.003,
\end{equation}
which is the dimensionless accretion rate for which the growth time is
equal to the Hubble time ($\sim$14\,Gyr). For significantly lower
accretion rates $\mdot \ll \mdotH$, the growth timescale is much
greater than the age of the Universe, and a given black hole cannot
possibly gain a significant fraction of its mass at that $\mdot$.

For a simple estimate of the accretion luminosity $L$, we have
followed \citet{ho02} and converted from B-band to bolometric
luminosities adopting a constant bolometric conversion
($L=c_{B}\,L_{B}$, $c_{B}=17$). However, the actual bolometric
conversion varies, ranging from $\sim11-17$ for bright AGN and quasars
\citep[e.g.][]{Elvis94}, to $\sim24$ \citep[e.g.][]{ho99b,ho00} for
the most low-luminosity RIAF/ADAF systems. For bright AGN and quasars,
the bolometric conversion $c_{B}$ appears to depend just on
luminosity, although this of course may instead reflect a dependence
on the accretion rate.  For objects with moderate to large accretion
rates, we adopt the luminosity-dependent bolometric conversions of
\citet{Marconi04}, based on observations of the quasar spectrum over a
wide range of wavelengths and as a function of luminosity
\citep[e.g.,][]{Elvis94,George98,VB01,Perola02,Telfer02,Ueda03,VBS03}.
This gives a B-band luminosity
\begin{equation}
\log{(L_{B})}=0.80-0.067\mathcal{L}+0.017\mathcal{L}^{2}-0.0023\mathcal{L}^{3},
\end{equation}
where $\mathcal{L} = \log{(L/L_{\sun})} - 12$.  The important point is
that $c_{B}$ {\em decreases} with increasing luminosity; i.e.\ the
brightest objects are the most dominated by the optical-UV portion of
the spectrum \citep[see also, e.g.][]{Wilkes94,Green95,Strateva05}.

At low luminosities, we consider the sub-sample of low-luminosity AGN
from \citet{ho99b} and \citet{ho00}, for which bolometric luminosities
from radio, optical, UV, and X-ray observations have been determined
directly. Calculating the $c_{B}$ correction for each of these
objects, we find no evidence for a direct correlation with $L$, but we
do find marginal evidence for a correlation of $c_{B}$ with observed
Eddington ratio $\lambda=L/\ledd$. We can fit this dependence roughly
to a power law and obtain the best fit
\begin{equation}
c_{B}\approx11\,(\lambda/0.01)^{-0.25}. 
\end{equation}
As essentially all these objects have $\mdot\lesssim\mdotcrit$
(defined below, but roughly $\mdotcrit\sim0.01-0.1$), we do not need
to worry about a ``break'' in the slope in terms of $\lambda$ induced
by the change in dependence of $\er$ on $\lambda$ (i.e.\ on $\mdot$)
at $\mdotcrit$. Again, we find that the ratio of total to B-band
luminosity decreases with increasing luminosity (though in this case
with dimensionless luminosity). This is also consistent with
theoretical spectral models of advection-dominated accretion
\citep[e.g.][]{Mahadevan97}, which generally feature an increasingly
hard X-ray spectrum and possibly increased contribution from jets at
low $\mdot$.

In order to determine which objects should receive which bolometric
correction, we divide our sample in two on the basis of our original
calculation of $\lambda$.  For objects with $\lambda>\mdotcrit$
(defined below), we apply the bolometric corrections of
\citet{Marconi04}, which are primarily derived from bright,
high-accretion rate objects. For objects with $\lambda<\mdotcrit$, we
apply the corrections derived from \citet{ho99b} and \citet{ho00},
whose samples are essentially all below this accretion rate. Although
the exact point at which we determine the divide in bolometric
corrections is somewhat arbitrary, this choice is consistent with
theory and observations in both cases, and it is not a significant
uncertainty as the two sets of corrections give similar results for
the objects in the transition region.

Observations of X-ray binaries have revealed a number of distinct
spectral states, such as the high soft state, low hard state,
quiescent state, etc. (see McClintock \& Remillard 2005 for a review).
Moreover, the same black hole can exhibit different states at
different times.  A primary cause of the various states is changes in
the mode of accretion, as suggested by Narayan (1996) and Esin,
McClintock \& Narayan (1997).  According to this model, accretion
flows with $\dot m$ greater than a critical value $\dot m_{\rm crit}$
are radiatively efficient and are well-described by a thin disk model.
These systems correspond to the high soft spectral state.  However,
for $\dot m < \dot m_{\rm crit}$, the accretion flow switches to a
two-zone state in which the thin disk is restricted to radii $R$
greater than a transition radius $R_{\rm trans}$, while the gas inside
$R_{\rm trans}$ accretes via a radiatively inefficient accretion flow
(RIAF), also known as an advection-dominated accretion flow (ADAF;
Narayan \& Yi 1994, 1995b, Narayan 2004).  This is the low hard state,
or (at very low $\dot m$) the quiescent state.  The transition radius
$R_{\rm trans}$ is only slightly larger than the radius of the
innermost stable circular orbit when $\dot m$ is just below $\dot
m_{\rm crit}$, but it increases to larger values as $\dot m$ decreases
further (see Yuan \& Narayan 2004).  Consequently, the radiative
efficiency is fairly close to $\epsilon_r^0=0.1$ when $\dot m$ is just
below $\dot m_{\rm crit}$, but it decreases rapidly (but smoothly) as
$\dot m$ decreases below $\dot m_{\rm crit}$ (see
Equation~[\ref{eqn:rad.eff}] below).

Observations suggest that the different accretion states observed in
X-ray binaries exist as well in AGN
\citep{NYM95,NYM96,Meier01,MGF03,XBONGS}.  It appears that the
transition between accretion states occurs at more or less the same
critical Eddington ratio $\mdot=\mdotcrit$, regardless of the mass of
the accreting black hole.  Although the value of $\dot m_{\rm crit}$
is somewhat uncertain (see the discussion below on hysteresis),
observations of black hole binaries \citep{Maccarone03} as well as
theoretical extensions of accretion models
\citep[e.g.,][]{NY95,EMN97,MLMH00} suggest $\mdotcrit\sim0.01$. Recent
observations of both a bimodal distribution of Eddington ratios at low
redshift \citep{Marchesini04} and the distribution of objects in the
$L-M$ plane \citep{Jester05} also suggest that the transition occurs
at $\mdot\sim0.01$ \citep[see][]{CX05}. Note that the similarity
between $\mdotcrit$ and $\mdotH$ is entirely coincidental.

In accretion models, the radiative efficiency does not depend on the
absolute value of $M$ or $\Mdot$, but only on the dimensionless
accretion rate $\mdot$ \citep[e.g.,][]{Chen95,EMN97,Abramowicz05}.
Based on observations and theoretical models, we model the radiative
efficiency $\er(\mdot)$ as
\begin{equation}
  \er = \left\{ \begin{array}{ll}
      \eo  & \mathrm{ if\ } \mdot > \mdotcrit \\
      \eo\,\Bigl( \frac{\mdot}{\mdotcrit}\Bigr) & \mathrm{ if\ } \mdot \leq \mdotcrit\ .
\end{array}
    \right.
\label{eqn:rad.eff}
\end{equation}
This particular choice for the efficiency factor was originally
suggested by early ADAF models \citep{NY95}, but it is generally
consistent with observations and ensures that the radiative efficiency
is continuous across the critical Eddington ratio $\mdotcrit$. With
this model, we can determine the accretion rate $\mdot$ for any black
hole given the bolometric luminosity $L$ and black hole mass $M$. The
model of $\er(\mdot)$ may actually overestimate the accretion rate for
a given luminosity at low $\mdot\ll\mdotcrit$.  This is because ADAFs
tend to suffer large mass loss through winds (Narayan \& Yi 1994,
1995a; Blandford \& Begelman 1999).  As a result, $\dot m$ decreases
with decreasing radius, so that the efficiency depends on whether one
considers the accretion rate at the black hole or at the transition
radius.  In terms of $\dot m$ at the black hole, which is the quantity
of interest for this paper, $\epsilon_r$ decreases less steeply with
decreasing $\dot m$, say as $\sim(\dot m/\dot m_{\rm crit})^{1/2}$;
i.e., the accretion rate for a given luminosity would be less than
Equation~(\ref{eqn:rad.eff}) predicts.  However, for our analysis we
choose the prescription given in Equation~(\ref{eqn:rad.eff}), which is
the conservative choice since it errs on the side of overestimating
the accretion rate at low $\lambda$.

What value of $\dot m_{\rm crit}$ should we use?  It has been observed
in X-ray binaries that there is a hysteresis phenomenon in the
transition between radiatively inefficient and radiatively efficient
accretion flows (see, e.g.\ Maccarone \& Coppi 2003 and references
therein; Barret \& Olive 2002). The transition from the radiatively
inefficient state (ADAF) to the radiatively efficient (thin-disk)
state often occurs at a fairly large accretion rate $\mdot \sim 0.1$
whereas the reverse transition (from radiatively efficient, to
inefficient accretion) occurs at a much lower $\mdot \sim 0.01$.  This
implies that when $\mdot$ {\em increases} with time, the critical
accretion rate may be as high as $\mdotcrit\sim0.1$, whereas when
$\mdot$ decreases with time $\mdot_{\rm crit}\sim0.01$.  Since the
peak iof quasar activity occurred at high redshift and the activity
declined more or less monotonically to the present epoch, one expects
that the lower value of $\mdotcrit\sim0.01$ is more relevant for our
analysis. However, the possibility of hysteresis does introduce a
degree of uncertainty in our model.  Therefore, we consider in our
subsequent analysis both $\mdotcrit=0.01$ and $\mdotcrit=0.1$, and
ultimately find that the choice does not qualitatively change our
conclusions.

\begin{figure*}
    \centering \plotone{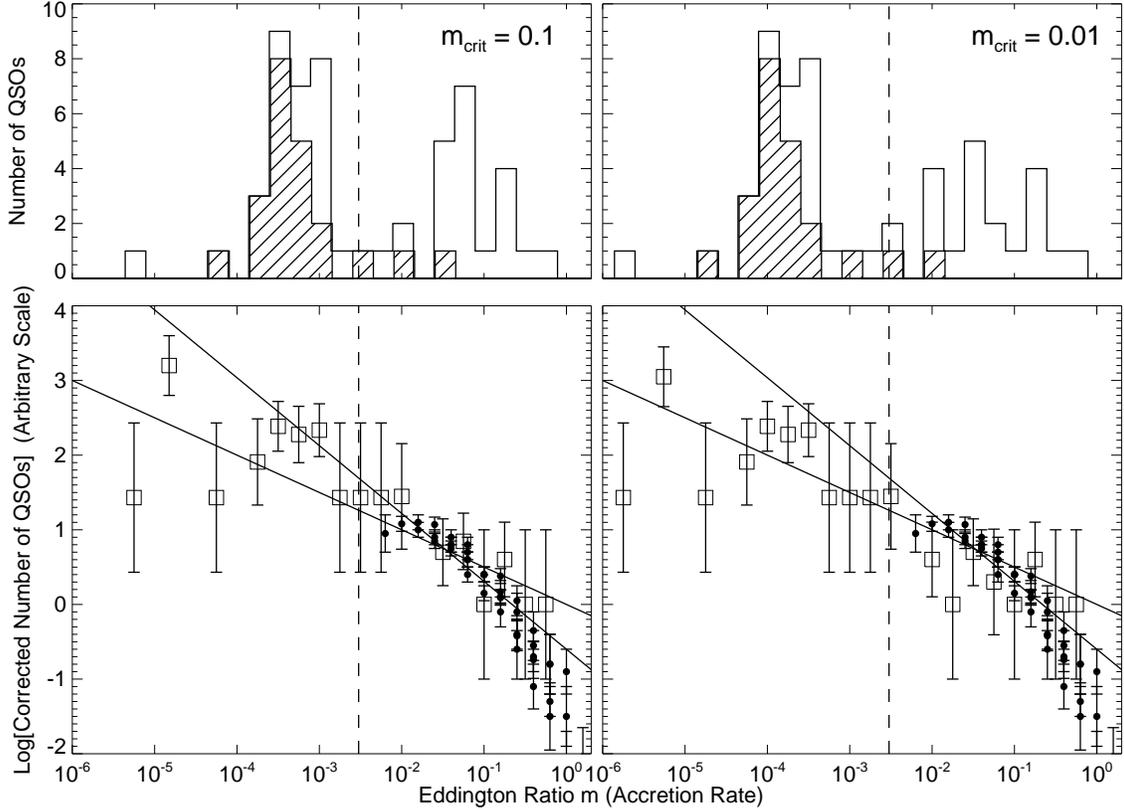} 
    \caption{Upper panels: The distribution of dimensionless accretion
    rate ($\mdot\equiv\Mdot/\Mdotedd$) calculated through
    eq. (\ref{eqn:rad.eff}) and binned by $\Delta\log\mdot=0.25$,
    from the sample of \citet{ho02}. Left panels assume
    $\mdotcrit=0.1$, right panels assume $\mdotcrit=0.01$. Shaded
    histograms show the distribution for AGN with elliptical host
    galaxies only.  Lower panels: Estimated relative number density of
    objects as a function of $\mdot$ (black points), 
    corrected using a simple visibility volume
    correction (alternatively, corrected to match the local AGN luminosity
    functions of \citet{Hao05}) for the combined samples of 
    \citet{ho02} and \citet{Marchesini04} (squares).
    Dots show the SDSS Eddington ratio distributions
    determined in \citet{YLK05} for systems with velocity dispersions
    $\sigma=70-110\,{\rm km\,s^{-1}}$.  The dashed line in all panels corresponds
    to the accretion rate below which the black hole growth timescale
    $M/\Mdot$ is larger than the age of the Universe.
    \label{fig:mdot.dist}}
\end{figure*}

Figure~\ref{fig:mdot.dist} shows the distribution of dimensionless
accretion rates $\mdot$ obtained from the observed sample using
Equation~(\ref{eqn:rad.eff}) (upper panels).  The left panel shows the
result for $\mdotcrit=0.1$, the right for $\mdotcrit=0.01$,
essentially bracketing the range allowed by the hysteresis effect
discussed above.  In each case, we show the distribution for all
objects (empty histograms) and for just those objects which are
morphologically identified as ellipticals (morphological type index
$T\leq-2$, shaded histograms).  We also show the value of $\mdotH$
(dashed line), the accretion rate for which the black hole growth
timescale is equal to the Hubble time.  Although the result is
slightly stronger for the elliptical subsample, the critical point is
the same in both cases: even after accounting for the fact that the
black holes are in a radiatively inefficient state and thus the actual
accretion rate is significantly larger than the observed luminosity
Eddington ratio, most local objects are accreting at Eddington ratios
$\mdot\ll\mdotH$; i.e.\ at very low accretion rates at which
significant fractional mass gain in less than the age of the Universe
is not possible. Although the choice of $\mdotcrit$ systematically
shifts the inferred $\mdot$ at very low $\mdot\ll\mdotcrit$, it has
almost no effect in the range of interest,
$\mdot\sim10^{-3}-10^{-1}$. The bolometric corrections we adopt also
slightly increase the characteristic $\mdot$ over that implied by
e.g.\ a constant $c_{B}=17$ (in other words, simpler bolometric
corrections only strengthen this point), although the difference is
negligible for all but the smallest ($\mdot\lesssim10^{-4}$) and
largest ($\mdot\gtrsim0.1$) accretion rates.

We can also attempt to use the observed sample accretion rate
distribution to (very roughly) infer that of the population as a
whole.  We estimate a simple visibility volume correction, using the
distances to each object in \citet{ho02}. Comparing the maximum
distances or distances within which $\gtrsim90\%$ of each (dynamical
or reverberation-mapping mass measurements) subsample are located,
this gives a rough ratio of the subsample effective volumes.
Alternatively, we correct the relative number density of sources in each $L$ bin to the
value implied by the $z\sim0$ low-luminosity SDSS AGN luminosity
function of \citet{Hao05}. In either case, the estimate is similar. Whether or 
not we include the \citet{Marchesini04} sample in this estimate also makes 
little difference. 
In the lower panels of Figure~\ref{fig:mdot.dist} we use this
correction (technically, the luminosity function correction 
including the \citet{Marchesini04} objects) to estimate a {\em
relative} $\mdot$ distribution (squares), with assumed Poisson
errors (logarithmically scaled). We caution that these
distributions are rough estimates and should be taken heuristically,
as in truth the selection effects involved in measuring individual
black hole masses are much more complicated than, e.g., a simple
magnitude limit.  We also show the observed SDSS Eddington ratio
distributions estimated in \citet{YLK05}, in a more robust statistical
sense, but for relatively high $\mdot\gtrsim10^{-3}-10^{-2}$. Again,
the normalization of these is essentially arbitrary, but the trend
with $\mdot$ should be robust.
We show these observations as the filled dots, for systems with
velocity dispersions $\sigma=70,\ 80,\ 90,\ 100,\ 110\,{\rm
km\,s^{-1}}$.

In all cases, the $\mdot$ distributions broadly agree, and suggest
that the $\mdot$ distribution can be approximated as a power-law (or
series of power laws) over a wide range in $\mdot$.  The solid lines
in the lower panels of Figure~\ref{fig:mdot.dist} show power-laws with
slopes of $-1/1.23$ estimated from \citet{YLK05} and $-1/2$ predicted
in \citet{H05e}, which give reasonable approximations over a wide
range of $\mdot$ and demonstrate the range in slope allowed. For a
power-law $\mdot$ down to a turnover at $\mdot_{\rm min}$, the
differential duty cycle takes the form
\begin{equation}
\frac{{\rm d}f}{{\rm d}\log{\mdot}}=(\beta\,\ln{10}\,\mdot_{\rm
min}^{\beta})\,\mdot^{-\beta}.
\label{eqn:pwrlaw.mdot}
\end{equation}
Essentially all of the observational uncertainty regarding duty
cycles, the proper effective volume and number density corrections for
these samples, and the depth at which the $\mdot$ distribution turns
over (as compared to e.g.\ selection incompleteness) is neatly
contained in the parameter $\mdot_{\rm min}$. The estimates from
observations in Figure~\ref{fig:mdot.dist} do at least provide a
reasonable upper limit $\mdot_{\rm min}\lesssim10^{-5}$, which allows
us to calculate the quantities of particular interest (e.g.\ upper
limits to the mass gain in low-efficiency phases), given that $\beta$
is also reasonably well-constrained in the range above.
Although these estimates are by no means a rigorous determination of
the $\mdot$ distribution, they do provide approximate constraints and
upper limits, which imply that our subsequent calculations (e.g.\ the
mass-weighted effective $\mdot$ and $\epsilon_{r}$) cannot be
qualitatively changed within the allowed uncertainty in
Figure~\ref{fig:mdot.dist}.

\subsection{Accretion Rates as a Function of Host Morphology}
\label{sec:morph}

\begin{figure}
    \epsscale{1.15}
    \centering \plotone{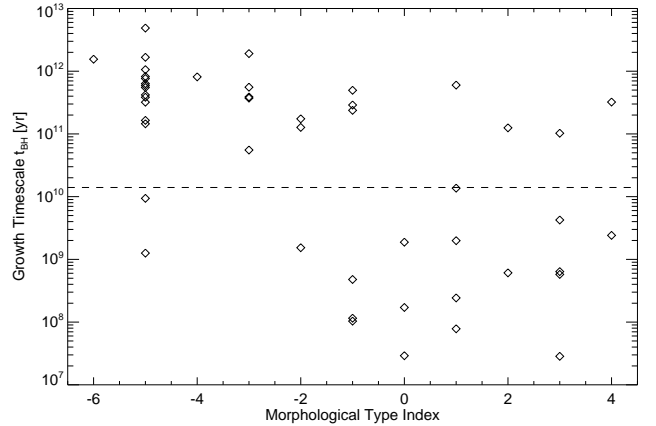} 
    \caption{The black hole growth timescale $\tBH=t_{S}/\mdot$ as a
    function of host galaxy morphology, for the sample from
    \citet{ho02}. Dashed line marks the Hubble time.
    \label{fig:morph}}
\end{figure}

Using the intrinsic mass accretion rates estimated above, accounting
for both the dependence of radiative efficiency on accretion rate and
spectral shape (bolometric corrections) on luminosity and accretion
rate, we can consider whether there is a difference in the
distribution of accretion rates in systems of different morphological
types.  In Figure~\ref{fig:morph}, we show the black hole growth
timescale $\tBH=t_{S}/\mdot$ for each object in the sample of
\citet{ho02} as a function of the host galaxy morphology
(morphological type index), where the host galaxy Hubble types are
determined in \citet{RC3} (RC3).  The dashed line in the figure shows
the Hubble time, $t_{\rm H}\approx14$\,Gyr.  It is clear that
essentially all objects with moderate to large accretion rates (i.e.\
$\tBH<t_{\rm H}$) are spirals, whereas almost every elliptical has a
negligible mass accretion rate ($\tBH\gg t_{\rm H}$).  This is also
clear in Figure~\ref{fig:mdot.dist}, and is well known from a number
of other observations \citep[see e.g.,][]{Heckman04}.

Observational studies of the $z=0$ black hole mass function estimated
from observations of galaxy luminosity functions and the distribution
of spheroid velocity dispersions have established that the total
present black hole mass density is dominated by black holes in
elliptical hosts, with at most a comparable contribution from black
holes in S0 host galaxies \citep[e.g.][]{AR02,Marconi04,Shankar04}.
If, as is clear in Figure~\ref{fig:morph}, the large majority of
present elliptical galaxies have $\tBH\gg t_{\rm H}$, this implies
that the majority of the present black hole mass density cannot have
been accumulated in the observed low-accretion rate phase, but that
ellipticals must have experienced a previous phase with
high-efficiency accretion.

Furthermore, regardless of whether or not all bright quasar activity
is the result of galaxy mergers or interactions, it is difficult to
imagine a major merger which does {\em not} excite bright quasar
activity (so long as a reasonable amount of cold,
rotationally supported gas is present in either
of the merging galaxies).  If elliptical galaxies are formed from the
merger of gas-rich spiral progenitors, as is expected in hierarchical
scenarios of structure formation (e.g.\ Toomre \& Toomre 1972, Toomre
1977; for a review, see Barnes \& Hernquist 1992), then it is
difficult to construct a scenario whereby an elliptical galaxy forms
without {\em some} period of rapid, high-$\mdot$ accretion, which need
only occur for $\gtrsim10^{6}$\,yr to dominate an entire Hubble time
worth of accretion at the present ($z=0$) rate.
Indeed, the gas
dissipation that would fuel the growth of supermassive black holes
is required if galaxy mergers are to explain
the high phase space densities (e.g. Hernquist et al. 1993), kinematic
properties (e.g. Cox et al. 2005), and fundamental scaling relations
(e.g. Robertson et al. 2005)
of ellipticals.


\section{Models for the Evolution of Black Hole Mass}
\label{sec:models}

\subsection{If Black Holes Accrete at Constant $\mdot$}
\label{sec:nohighedd}


\begin{figure}
    \epsscale{1.15}
    \centering \plotone{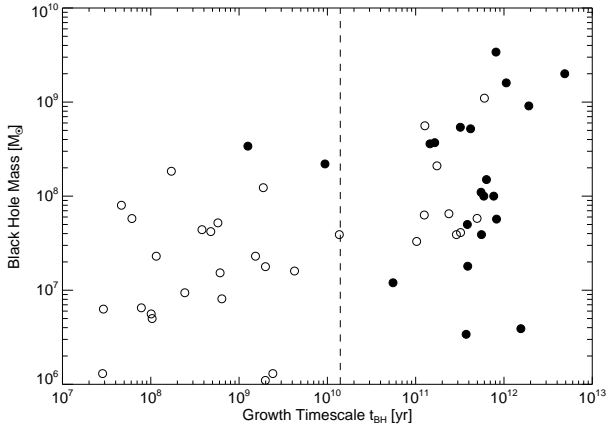} 
    \caption{The black hole growth timescale as a function of black
    hole mass, open circles the entire sample and filled
    ellipticals. Dashed line shows the Hubble time.
    \label{fig:dM.dist}}
\end{figure}

Could each black hole accrete at roughly a constant $\mdot$ during its
entire lifetime?  If this is the case, then objects with
$\mdot\ll\mdotH=0.003$ have $\tBH\gg t_{H}$, and do not gain mass in a
Hubble time, whereas objects with $\mdot\gtrsim10^{-2}$ with $\tBH\ll
t_{H}$ experience many $e$-foldings. Thus there should be a strong
change in the black hole mass distribution at $\tBH=t_{H}$, with much
higher black hole masses for $\tBH<t_{H}$.  Figure~\ref{fig:dM.dist}
shows the black hole mass as a function of growth time $\tBH$ for each
object in our sample (ellipticals filled, late-types open), with the
dashed line marking the Hubble time. Clearly, the predicted trend is
not evident; if anything, the most massive objects in our sample have
the {\em lowest} accretion rates (highest $\tBH$).  This trend has
been studied in greater statistical detail by \citet{Heckman04}.


\subsection{Models Involving Time Evolution of $\mdot$}

The conclusion of the previous subsection is that $\mdot$ must vary
with time and that the mass growth in many systems must occur in an
earlier phase of high $\mdot$ accretion.  We now ask whether low
efficiency, low accretion rate growth can give a significant
fractional contribution to the absolute black hole mass density or
individual black hole masses at $z=0$. The answer to this depends on
the time history of the accretion rate, for which we need to adopt
some reasonable model.

Consider a general model in which $\mdot=1$ at early times and then
decays as a power-law in time, peak phase with $\mdot=1$,
\begin{equation}
\mdot=1/(1+(t/t_{Q})^{\eta}).
\label{eqn:mdot.decay}
\end{equation}
Any such model will naturally produce a power-law distribution of
accretion rates at low $\mdot$ (Equation~[\ref{eqn:pwrlaw.mdot}]) with
$\beta=1/\eta$ and $\mdot_{\rm min}=(t_{H}/t_{Q})^{-\eta}$ (and in the
appropriate limits, can also represent exponential or step-function
decay).  The observed $\mdot$ distributions are broadly consistent
with the above model for $\eta\sim1.25-2.0$ (see
Figure~\ref{fig:mdot.dist}).  Simulations of quasar light curves
\citep{H05g} suggest $\eta\approx2$, as does connecting $z=0$
accretion rates at $t\sim t_{H}$ to a bright $\mdot\sim1$ phase of
duration $\sim10^{7}-10^{8}\,$yr implied by a wide range of
observations (see Martini 2004, and references therein).  We therefore
adopt $\eta=2$ (this choice also allows analytical solutions below),
but our calculations are only changed by factors $\sim\eta/2$ for
other values of $\eta$. Simulations and theoretical models
\citep[e.g.][]{SR98,Fabian99,CO01,WL03,DiMatteo03,DiMatteo04,DSH05}, 
a wide range of
observations \citep[e.g.,][]{Martini04}, and our estimated upper limit
$\mdot_{\rm min}\lesssim10^{-5}$ (simply from the existence of a
significant population of low-$\mdot$ observed ellipticals) imply a
bright (high-$\mdot$) quasar lifetime $\sim10^{6}-10^{8}\,$yr, setting
a fairly strict upper limit $t_{Q}\lesssim
t_{S}\sim4\times10^{7}\,$yr.

The above model of $\mdot(t)$ with $\eta=2$ gives, for an initial
black hole mass $M_{0}$, the following result for the mass growth as a
function of time,
\begin{equation}
M(t)=M_{0}\,\exp{\Bigl\{}\frac{t_{Q}}{t_{S}}\,\tan^{-1}\left({t\over
t_{Q}}\right){\Bigr\}} \, ;
\end{equation}
i.e.\ the object grows exponentially as $M=M_{0}\exp(t/t_{S})$ for
$t\ll t_{Q}$ and asymptotes to
$M_{f}=M_{0}\,e^{(\pi/2)\,(t_{Q}/t_{S})}$ as $t\rightarrow\infty$.  If
$t_{Q}\gg t_{S}$, then growth is large, but this is because the black
hole spends a great deal of time in bright, high-$\mdot$ phases before
the accretion rate slowly drops off. In fact, even after a Hubble
time, the accretion rate would still be large.
To reach $\mdot\leq0.01$ by $t=t_{H}$ requires
$t_{Q}\lesssim3\,t_{S}$, yielding a small mass gain at low $\mdot$;
the maximum fraction of the present black hole mass which can be
accreted at $\mdot<0.01$ by the present is $\sim8\%$ with
$t_{Q}\approx1.8\,t_{S}$.

More complex and host-galaxy dependent quasar light curves have been
obtained in hydrodynamical simulations of galaxy mergers incorporating
gas cooling, a multiphase interstellar medium, Bondi-Hoyle accretion
determined from the surrounding gas properties, and thermal feedback
from accretion \citep{SH02,SH03,SDH05b}. Despite the fact that these
light curves show most objects spend significantly more time at
luminosities and accretion rates well below the peak quasar
luminosity \citep{H05c,H05d,H05e}, the authors find that the mass
growth is dominated by efficient phases, with $\gtrsim70-80\%$ of mass
accumulated in bright, optically observable $\mdot\sim1$ quasar phases
(rising to $\sim100\%$ of the mass in the most massive objects
$M\gtrsim10^{9}\,M_{\sun}$), and the remaining mass essentially
entirely acquired in the short following phase of declining $\mdot$ at
moderate $\mdot\sim0.1$ \citep{H05e}.

\subsection{The Mass-Weighted Eddington Ratio and Radiative Efficiency}

Using the parameterization of the quasar light curve given in
Equation~(\ref{eqn:mdot.decay}), we can define an ``effective''
mass-weighted accretion rate $\EV{\mdot}$, i.e.\
\begin{equation}
\EV{\mdot}=\frac{1}{M_{f}}\int{\mdot(t)}\,\dot{M}\,{\rm d}t,
\label{eqn:mdot.wt}
\end{equation}
which can be evaluated numerically for a given $t_{Q}/t_{S}$. This
gives $\EV{\mdot}\approx0.50,\ 0.34$ for $t_{Q}=10^{7}\,$yr and
$t_{Q}=t_{S}$ respectively, in good agreement with other estimates of
the typical accretion rates at which black holes gain most of their
mass \citep[e.g.,][]{AGES}. It requires a very large
$t_{Q}\gtrsim10\,t_{S}$ to give low $\EV{\mdot}\lesssim0.02$, but such
long quasar lifetimes are ruled out by a number of arguments as
discussed above.  Using the fitted power laws of \citet{YLK05} gives a
significantly shallower decay, $\eta\approx1.23$, but similar
$\EV{\mdot}\approx0.21,\ 0.14$ for $t_{Q}=10^{7}\,$yr and
$t_{Q}=t_{S}$. If we use the binned $\mdot$ distribution of 
Figure~\ref{fig:mdot.dist} directly, we obtain $\EV{\mdot}\approx0.3,\ 0.2$ for
$t_{Q}=10^{7}\,$yr and $t_{Q}=t_{S}$, with a negligible difference if
we assume $\mdotcrit=0.01$ or $\mdotcrit=0.1$. The difference due to how 
we estimate the correction to an ``intrinsic'' $\mdot$ distribution is also small 
compared to the expected Poisson noise from the small sample. 
Assuming the 
distribution cuts off at the lowest observed $\mdot$ 
sets an upper limit on the duty cycle (lower limit on $\EV{\mdot}$) 
independent of $t_{Q}$ (see Equation~\ref{eqn:pwrlaw.mdot}),
giving $\EV{\mdot}\gtrsim0.18$.

We can similarly define an ``effective'' mass or luminosity-weighted
radiative efficiency, replacing $\mdot$ with $\epsilon_{r}$ in
Equation~(\ref{eqn:mdot.wt}) above.
Our ``canonical'' radiative efficiency $\epsilon_{r}^{0}=0.1$ can be
factored out of this equation, and we can then numerically calculate
the resulting $\EV{\epsilon_{r}}$ for a given $t_{Q}/t_{S}$ and
$\mdotcrit$.
For a relatively large $t_{Q}=t_{S}$, we obtain
$\EV{\epsilon_{r}}=0.81\,\epsilon_{r}^{0}\ (\mdotcrit=0.1);\
0.93\,\epsilon_{r}^{0}\ (\mdotcrit=0.01)$. For a shorter timescale
$t_{Q}\sim10^{7}\,$yr these estimates increase to
$0.95\,\epsilon_{r}^{0}\ ,\ 0.98\,\epsilon_{r}^{0}$.
Likewise, if we adopt just the observed distribution of accretion
rates as the PDF for accretion rate over each AGN lifetime as for
$\EV{\mdot}$, we obtain $\EV{\epsilon_{r}}=0.92\,\epsilon_{r}^{0}\
(\mdotcrit=0.1);\ 0.96\,\epsilon_{r}^{0}\ (\mdotcrit=0.01)$ for
$t_{Q}\sim 10^{7}\,$yr, and
$\EV{\epsilon_{r}}=0.72\,\epsilon_{r}^{0}\ (\mdotcrit=0.1);\
0.85\,\epsilon_{r}^{0}\ (\mdotcrit=0.01)$ for $t_{Q}\sim t_{S}$ (and
values closer to $1.0\epsilon_{r}^{0}$ for $t_{Q}<t_{S}$).  If we were
to weight by luminosity instead of mass, we would be even further
biased towards $\EV{\epsilon_{r}}\approx\epsilon_{r}^{0}$.

Thus, for reasonable values of the quasar lifetime, the ``effective''
radiative efficiencies expected are quite similar to the canonical
values adopted for bright quasars; essentially, this is a restatement
of our previous derivation that most mass/luminosity is
accumulated/radiated in radiatively efficient, high accretion rate
phases.


\subsection{Comparison with Expected Quiescent Accretion Rates}

We can also attempt to estimate the accretion rates and growth
timescales of the elliptical galaxies observed. If we assume that the
gas in these galaxies is virialized and is in spherical hydrostatic
equilibrium and adopt a \citet{Hernquist90} profile for the static
potential (set by stars and dark matter) we can determine the central
sound speed and density, and estimate the accretion rate at the
Bondi-Hoyle rate. Using the $M-\sigma$ and $M-M_{\rm sph}$ relations
measured by \citet{Tremaine02} and \citet{MH03}, respectively, we
obtain
\begin{equation}
\begin{split}
\tBH=&\alpha\frac{1}{f_{\rm gas}}\frac{M_{\rm sph}}{M_{\rm BH}}\,t_{\rm dyn}\\
&\sim10^{12}\,{\rm yr}\ \alpha\,\Bigl( \frac{0.01}{f_{\rm gas}}\Bigr)\,
\Bigl( \frac{M_{M-\sigma}}{M}\Bigr)\,\Bigl( \frac{\sigma}{200\,{\rm km\,s^{-1}}}\Bigr),
\end{split}
\label{eqn:Bondi}
\end{equation}
where $\alpha\sim1$ is a constant which depends on the profile and gas
equation of state, $\approx0.6$ for the assumptions above and
$\gamma=5/3$, $t_{\rm dyn}$ is the spheroid dynamical time, $f_{\rm
gas}$ is the gas mass fraction, and $M_{M-\sigma}$ is the expected
$M-\sigma$ black hole mass.

Comparison with Figure~\ref{fig:dM.dist} suggests that
Equation~(\ref{eqn:Bondi}) provides a good order-of-magnitude estimate
for the ellipticals in our observed sample (typical
$M\sim10^{8}\,M_{\sun}$). Of course, there is no particular evidence
in the observations for the weak dependence on $\sigma$ given by our
model. But this further suggests that these objects must have gone
through a previous bright phase, to heat the gas to virial
temperatures and lower the gas fraction to $\sim1\%$ typical in
observed ellipticals. Moreover, note that $\tBH\propto
M_{M-\sigma}/M$; i.e.\ if a black hole is undermassive, it will take
longer to grow, meaning it is unlikely that these black holes grew to
their observed $M-\sigma$ masses via slow, quiescent accretion.

Finally, in essentially all cases where the Bondi accretion rates of
low-$\mdot$ AGN have been determined
\citep{FC88,DiMatteo00,DCF01,Loewenstein01,Bower03,Pellegrini05},
as well as the same in X-ray binaries \citep[e.g.][]{Perna03}, it is
observed that objects are actually accreting well {\em below} the
Bondi rate. Thus, the fact that our estimate above provides an
approximate agreement with the observed growth timescale suggests that
we may actually be significantly overestimating the accretion rates of
low-$\mdot$ AGN, making our conclusions even stronger.  This
may be related, for example, to the fact that in the Milky Way case,
an external observer would likely associate a significant amount of
nearby but (owing to our proximity compared to AGN in other galaxies)
separately resolved emission with Sgr A$^{\ast}$, because the AGN
luminosity is not large enough to overwhelm these sources 
\citep[see the discussion in][]{ho02}. 

\section{Conclusions}
\label{sec:conclusions}

We have determined the distribution of accretion rates of $z=0$
supermassive black holes from the sample of \citet{ho02}, including
the effects of changing radiative efficiency at low accretion rates in
radiatively inefficient accretion flows and the dependence of
bolometric corrections on both luminosity and accretion rate. We find
that most local black holes have very low accretion rates
$\mdot\sim10^{-4}$, well below the minimum accretion rate
$\mdotH=0.003$ required to grow significantly over the entire age of
the Universe. This is especially true in the case of supermassive
black holes in elliptical galaxies, implying that the dominant part of
the integrated black hole mass density was formed in radiatively
efficient high-$\mdot$ phases, as is expected if these galaxies have
undergone mergers in the past which would fuel quasar activity and
eventually heat or expel gas in some form of self-regulated growth
(Di Matteo et al 2005, Springel et al. 2005a).

We have further discussed models for the accretion rate as a function
of time or distribution of accretion rates given the observed $z=0$
distribution, and find in all cases that the accumulated mass should
be dominated by mass gained in efficient, high-$\mdot$ phases.  It is
only possible to gain significant mass in low-$\mdot$ phases in these
models with a high degree of fine-tuning of the quasar lifetime,
giving values inconsistent with both theory and
observations. Furthermore, scaling with redshift is not expected to
change our results significantly, and it is increasingly difficult for
black holes to gain any mass in low-$\mdot$ states if a previous,
high-efficiency mass gain period is suppressed. 

We do not expect that selection effects will significantly change our
results. We have compared directly the sample of \citet{ho02} and the
combined samples of \citet{ho02} and \citet{Marchesini04}, and find
identical qualitative results.  We have considered separately
different classes of black holes measured by different techniques,
namely reverberation-mapping and stellar kinematics, and find that the
two methods yield similar qualitative results despite having
essentially opposite biases.  However, because the two methods have
opposite biases, current measurements with either method only
minimally overlap at bolometric luminosities $L\sim10^{9}\,L_{\sun}$,
allowing the possibility of a bias against sources with accretion
rates $\mdot\sim10^{3}-10^{-1}$. The addition of the
\citet{Marchesini04} sample spans this gap in luminosity, while
preserving the deficit of objects at these accretion rates.

Based on the $z=0$ distribution of accretion rates, the integrated
mass of the black hole population must have been gained in earlier,
radiatively efficient, high-$\mdot$ ($\mdot\gtrsim0.1$) phases.
Likewise, the mass growth of most individual black holes must be
dominated by high-$\mdot$ rapid growth.  This is in good agreement
with a range of inferred accretion histories from observations of
quasar populations
\citep[e.g.][]{Soltan82,Salucci99,YT02,Marconi04,Shankar04}, and
direct observations of quasar accretion rates and their evolution with
redshift \citep{Vestergaard04,Heckman04,MD04,YLK05}.  It similarly
agrees well with theoretical models of the relation between black hole
mass and host galaxy mass or velocity dispersion
\citep{SR98,Fabian99,CO01,WL03,DSH05} and semi-analytical modeling of
galaxy and quasar formation \citep[e.g.][]{KH00}. Recently, more
detailed modeling of quasar light curves and the implied rate of
formation of sources from the observed luminosity functions (Hopkins
et al.\ 2005a-g) also predicts the black hole mass function allowing
for luminosity and host galaxy-dependent quasar lifetimes in which
mass growth is dominated by a short peak accretion phase of high
$\mdot$. Unlike all of these previous observational and theoretical
constraints, however, our present result is model-independent, and
does not involve large uncertainties in e.g.\ the evolution of quasars
and their fueling mechanisms.

Because the radiative efficiency of black hole accretion is expected
and observed to decrease at low $\mdot$, this argument is amplified
when considering the radiative luminosity of accreting black holes.
That is, the radiative output is even more dominated by high-$\mdot$
phases of accretion.  This is consistent with the cosmic IR or X-ray
backgrounds, synthesis models of which suggest black hole growth
dominated by high-$\mdot$ phases
\citep[e.g.][]{Comastri95,Gilli99,Ueda03,H05e} and relatively short
time spent at $\mdot\sim\mdotcrit$ \citep{Cao05}, and also with simple
arguments which suggest radiative efficiencies equal to or even larger
than $\epsilon_{r}=0.1$ \citep{ERZ02,Merloni04a}.  Thus, models in
which quasars spend a very large fraction of their time at low
luminosities or low accretion rates (e.g.\ Hopkins et al.\ 2005a-g)
are completely consistent with both the observed $z=0$ distribution of
accretion rates and the above observations which all suggest that mass
growth and radiative output are dominated by short, high-$\mdot$
phases.

Finally, we have not ruled out the possibility that some small
fraction of {\em individual} objects may gain significant mass in
low-$\mdot$ states. Based on the observed $\mdot$ distribution, this
fraction is small, $\lesssim5\%$, and we have shown that it does not
contribute significantly to the integrated black hole mass density.
However, it could have important implications for the energetics of
these objects and their relations to their host galaxies, as feedback
may couple differently to the surrounding medium (in, for example, the
form of jets instead of integrated disk radiation, given the
low-accretion rate state; e.g.\ Narayan \& Yi 1995a).  Likewise, it is
not clear whether the physical mechanisms responsible for the
$M-\sigma$ relation in most systems would apply (or whether
alternative mechanisms might have the same effect) in low-$\mdot$
systems (although see e.g.\ Churazov et al.\ 2005).

To summarize, the observed $z=0$ distribution of black hole accretion
rates constrains the total time black holes spend in different states
and the relative contribution of these states to the integrated final
black hole mass and luminosity output.

{\bf (1)} Supermassive black holes spend most {\em time} in a
low-efficiency, low accretion rate state. The $z=0$ Eddington ratio
distribution implies that the bright quasar lifetime $t_{Q}\lesssim
t_{S}$.  This means that $\sim1\%$ of a supermassive black hole's
lifetime is spent in a bright, high-$\mdot$ quasar phase and $\sim
99\%$ is spent in a dim low-$\mdot$ phase.

{\bf (2)} Nevertheless, the growth of black hole {\em mass} is
dominated by the short, high-$\mdot$ phase of evolution.
The Eddington ratio distribution implies an effective mass-weighted
$\EV{\mdot}\sim0.3$, in good agreement with estimates from a wide
range of other observations of quasars.

{\bf (3)} Likewise, the integrated {\em luminosity} radiated is
dominated by the high-$\mdot$, high radiative efficiency growth
phase. The effective luminosity-weighted radiative efficiency implied
by the observations, coupled with simple models of the quasar
evolution, is $\EV{\epsilon_{r}}\gtrsim 0.8-0.9\,\epsilon_{r}^{0}$,
where $\epsilon_{r}^{0}\sim0.1$ is the radiative efficiency at
$\mdot=1$. Again, this is consistent with a range of constraints from
observations.

We thus confirm current wisdom regarding points {\bf (1)} and {\bf
(3)}, and we obtain a robust result in the case of {\bf (2)}.

\acknowledgments We thank an anonymous referee for comments that 
significantly improved and clarified the text. 
This work was supported in part by NSF grants ACI
96-19019, AST 00-71019, AST 02-06299, AST 03-07433 and AST 03-07690,
and NASA ATP grants NAG5-12140, NAG5-13292, and NAG5-13381.

\end{document}